# Erbium-doped WS$_2$ with Down- and Up-Conversion Photoluminescence Integrated on Silicon for Heterojunction Infrared Photodetection

*Qiuguo Li,* [1] *Hao Rao,* [2] *Haijuan Mei,* [1] *Zhengting Zhao,* [1] *Weiping Gong,\*,* [1] *Andrea Camposeo,* [3] *Dario Pisignano,* [3, 4] *and Xianguang Yang\*,* [2]

[1]Guangdong Provincial Key Laboratory of Electronic Functional Materials and Devices, Huizhou University, Huizhou 516001, Guangdong, China

[2]Institute of Nanophotonics, Jinan University, Guangzhou 511443, China

[3]NEST, Istituto Nanoscienze-CNR and Scuola Normale Superiore, Piazza S. Silvestro 12, I-56127 Pisa, Italy

[4]Dipartimento di Fisica, Università di Pisa, Largo B. Pontecorvo 3, I-56127 Pisa, Italy

*E-mail: xianguang@jnu.edu.cn and gwp@hzu.edu.cn

# ABSTRACT

The integration of 2D nanomaterials with silicon is expected to enrich the applications of 2D functional nanomaterials and to pave the way for next-generation, nanoscale optoelectronics with enhanced performances. Herein, a strategy for rare earth element doping has been utilized for the synthesis of 2D $WS_2$:Er nanosheets to achieve up-conversion and down-conversion emission ranging from visible to the near-infrared region. Moreover, the potential integration of the synthesized 2D nanosheets in silicon platforms is demonstrated by the realization of an infrared photodetector based on a $WS_2$:Er/Si heterojunction. These devices operate at room temperature and show a high photoresponsivity of ~39.8 mA/W (at 980 nm) and a detectivity of $2.79 \times 10^{10}$ cm·Hz$^{1/2}$/W. Moreover, the dark current and noise power density are suppressed effectively by van der Waals assisted p-n heterojunction. This work fundamentally contributes to establishing infrared detection by rare element doping of 2D materials in heterojunctions with Si, at the forefront of infrared 2D materials-based photonics.

# 1. INTRODUCTION

Photodetectors for the infrared region have wide applications in many fields, such as remote surveillance, [1, 2] medical imaging, [3-5] and thermal imaging.[6, 7] Traditional infrared detectors focus on Si material [8] and HgCdTe [9] (<0.1 eV) materials due to their small bandgap for infrared light absorption. However, these materials-based infrared detector might suffer from low photoresponsivity, and from the need of cryogenic cooling, which is expensive, awkward and time-consuming. Recently, two-dimensional materials (2DMs) with van der Waals (vdW) interface bonds, that can be integrated with both silicon-based platform and soft materials, have been shown to have unique and highly promising properties as sensing elements of photodetectors.[10-13] Graphene with zero bandgap, which is compatible with other element doping throughout the fabrication of photodetectors, has been exploited in various devices and for high speed optical detection at telecom wavelengths.[14] It is highly desirable to fabricate other 2DM nanostructures with controllable morphology and tunable bandgap, to overcome issues arising from limited photoresponsivity of graphene.[15] Tungsten disulfide ($WS_2$) is one of the 2DMs exhibiting bandgap tunability, from 1.4 eV in bulk to 2.1 eV in monolayers.[16] Compared to the most investigated 2D material such as $MoS_2$[16], $WS_2$ has also been proven to be more chemically stable due to oxidation-against and resistant of high-rate thermal decomposition. Moreover, $WS_2$ shows superior electron mobility with respect to other 2DMs (of the order of $10^3$ $cm^2$ $V^{-1}$ $s^{-1}$),[17] as well as high absorption efficiency ($10^5-10^6$ $cm^{-1}$),[18] which are relevant properties for building blocks of photodetectors. With these appealing properties, 2D $WS_2$ layers are fully exploited as a fundamental part in assembly of high-performance optoelectronic devices, such as photodetectors,[19, 20] light harvesting for hydrogen production[21, 22] and batteries anode[23-25]. However, the surface defects in $WS_2$ nanosheets might promote enhanced local recombination and increase of the dark current, thus leading to low photoresponsivity and slow response time.[19] Proper material design and synthesis can be critically important to circumvent such limits. Among various strategies reported in recent years, the Si/2DM photodetectors [26-29] have gained much attention mainly due to the following two reasons. The heterojunction photodetectors are explored to suppress the dark current and improve the photoresponsivity of the device performance. In addition, they can facilitate the electron-hole separation and achieve response time of the order of microseconds.[30-32] Importantly, Si is a fundamental semiconductor material, it is widely used in high-performance, low-cost photonic integrated circuits and it is highly compatible with modern fabrication processes.[33, 34] However, there are a few challenges to the fabrication of 2D $WS_2$/Si heterojunction, which could be listed as follows. First, due to the large intrinsic bandgap of $WS_2$ [35-37] and the indirect bandgap of Si,[38] strong interface recombination at the heterojunction device would hurdle the improvement of device performance. Second, the fabrication of $WS_2$ nanosheets involves high cost, complex and energy-consuming process, along with structural limitation of the materials.[39-43] Besides, previous reports focus on $WS_2$ nanosheets by wet transferring to Si substrate for device fabrication, which bring unexpected and unavoidable damage to the device, and possibly decrease performance.[44,45] Many of these drawbacks can be circumvented by introducing defects within the bandgap of $WS_2$ nanosheets by rare-earth element substitution,[46] an approach that would be appealing also for applications for infrared photodetectors.[47] Moreover, the rare-earth fine structured spectral emission and absorption[48] offer many opportunities for fundamental research and technical applications of 2DM-based infrared photodetectors, due to the breakthrough of the intrinsic bandgap limitation. In addition, the rare-earth element can convert the infrared light into visible light, resulting in high photoresponsivity of detectors performance. Rare-earth ions doped in semiconductor hosts have abundant excited energy levels ranging from UV to infrared region due to unique intra 4f electronic transitions. Besides, high quantum yield, narrow bandwidth, long lived emission, high photostability of rare-earth ions in photoluminance characteristic are manifested. Thus, the photodetection activities achieved via coupling rare-earth ion ($Er^{3+}$, $Eu^{3+}$, $Tb^{3+}$) [47, 49-51] with semiconductors have been demonstrated significantly. Herein, infrared detector based on Er doped $WS_2$ materials focus the defect engineering and alleviate the strong interface recombination at the $WS_2$/Si heterostructure, resulting more photoexcited electrons transferring into the electrodes for collection with improved photodetectivity, which is an attractive strategy to harvest infrared light in two-dimensional materials.

In this paper, we report on the fabrication of erbium-doped $WS_2$/Si ($WS_2$:Er/Si) heterojunction infrared photodetectors by a simple two-step process, which include magnetically co-sputtering Er and W for preparing W:Er film on Si substrate, followed by sulfurization to $WS_2$:Er nanosheets/Si heterojunction. We found that $WS_2$:Er can form highly ordered layered structures on Si substrate. The Er doping not only increases the infrared light absorption, but also enhances the photodetectivity. The photodetectivity of the $WS_2$:Er nanosheets/Si structure is ~40 mA/W, with a response time of ~170 ms higher than that of the pure $WS_2$ nanosheets (0.8 mA/W) under the same test condition. More importantly, an improved stability is achieved because of the rational structural arrangement and enhanced photoelectrical properties, indicating that the $WS_2$:Er/Si infrared photodetectors are relevant for potential applications in infrared photodetectors.

## 2. EXPERIMETNAL SECTION

**2.1. Preparation of WS$_2$:Er/Si.** The heterojunctions were grown by a two-step procedure. First, the Er-doped W film was pre-deposited by co-sputtering of W and Er targets. Specifically, a 10 mm × 10 mm Si substrate, which serves as a growth substrate, was taken into a magnetic sputtering furnace after anhydrous ethanol clean and sonication. W target was provided as source with DC sputtering while Er target was used for doping under magnetic sputtering, since the deposition rate of magnetic sputtering is low. The distance between targets and substrate was ~5 cm. The carrier gas Ar was used as working gas to ignite the targets with pressure in the furnace of 0.74 Pa. The power for sputtering was 30 W for Tungsten and 10 W for Er, respectively, while the temperature of the substrate was kept at 300°C for the whole sputtering process. After 50 min of co-sputtering of W and Er targets, the Er-doped W film was removed. Next, an alumina boat containing S powder (~20 mg, 99.99%, from Aladdin) for sulfurization growth of WS$_2$:Er was placed at the upstream in a chemical vapor deposition (CVD) furnace, whereas the pre-deposited Er-doped W films was placed up-side in another alumina boat ~10 cm away from the S powder. The furnace was evacuated to a residual pressure of 0.1 mbar and then filled with Ar to remove the residual oxygen. The temperature of the furnace was increased up to 680 °C from room temperature in 85 minutes and then held at this temperature for 40 minutes. During growth, 120 sccm (standard state cubic centimeter per minute) of Ar was used as transporting gas to flush the furnace. After 40 min, the furnace was cooled down to room temperature under Ar gas flow and the samples were taken out for characterization.

**2.2. Material Characterizations.** A Zeiss scanning electron microscope (SEM) operating at an accelerating voltage of 5 kV was used to characterize the morphology of the WS$_2$:Er/Si heterojunction. Chemical compositions of the heterojunction were examined by energy dispersive X-ray spectroscopy (EDS) in the Zeiss SEM equipped with a Bruker AXS Quantax system working at 20 keV. X-ray diffraction (XRD, X′ Pert PRO MPD) was used to measure the crystallinity of the prepared heterojunction. The surface valence states of the elements in the $WS_2$:Er nanosheets were investigated by X-ray photoelectron spectroscopy (XPS, PHI-QUANTERA-2). Transmission electron microscope (TEM) and high-resolution TEM (HRTEM) images of the samples were taken with a Tecnai-G2-F30 field emission TEM operating at an accelerating voltage of 200 kV.

**2.3. Raman and PL Measurements.** Raman measurements of the $WS_2$:Er/Si samples were performed with a laser confocal microscope (HORIBA, XploRA), by utilizing a 50× objective (NA = 0.65) to focus the excitation laser onto the active area with a spot of ~1.0 μm in diameter. Raman measurements were performed under backscattering geometry with 532 nm laser excitation. The spectra were recorded from 20 to 1000 $cm^{-1}$ using an ultralow frequency filter, 1800 grooves/mm grating, and a Peltier cooled charge coupled device (CCD) with 150 s of acquisition time. Down- and up-conversion PL measurements were performed with 532-nm and 980-nm laser excitation, respectively, under different excitation intensities.

**2.4 Device Fabrication and Characterization.** To fabricate the $WS_2$:Er/Si heterojunction device, silver (Ag) plates acting as conductive pads were deposited with a blade knife onto the p-type Si (resistivity 0.1−5 Ω·cm) and $WS_2$:Er for negative and positive electrode, respectively, with the size of ~ 0.1 cm × 0.1 cm. The electrical and photoelectrical properties of the $WS_2$:Er/Si heterojunction device were characterized by a semiconductor characterization system (Keithley 4200-SCS, Tektronix), illuminated with 980-nm light sources. The illuminated area was determined by the hole area of a mask (∼0.2 cm$^2$) positioned onto the devices in photoelectrical measurements. The light that illuminated the photodetector could be switched on and off by an acousto-optic modulator, with a rise and fall time of 0.1 s, respectively.

**2.5. Numerical Simulation.** The electrical field distributions together with the I-V characteristics at dark conditions and under illumination were obtained by using a commercial electromagnetic software (COMSOL) with finite element (FE) solutions. The optical field distribution was also simulated by FE solution, and conduction band energy profiles of the device under light illumination were performed to study the enhancement mechanism of photoresponsivity.

**2.6 Statistical Analysis.** The data presented in the graphs of the paper are the average results of at least three independent measurements. Statistical analysis was carried out using Origin Software.

## RESULTS AND DISCUSSION

A large area of uniform WS$_2$:Er nanosheets grown on a Si substrate (see Figure S1, Supporting Information), collected after the simple two-step synthesis process described above, is also shown in Figure 1a. A dense ensemble of nanosheets is present on the surface of the Si substrate. The magnified SEM image displayed in Figure 1b evidence nanosheets with side length of 0.1−0.5 μm and thickness of ∼10 nm, and clearly show the layer structure. The composition of as-prepared WS$_2$:Er nanosheets is analyzed from EDS spectra. W, S and Er signals are observed (Figure 1c), which confirms the elemental contents of Er (atomic%, 0.86%), W (atomic%, 13.29%) and S (atomic%, 32.31%), giving the atomic ratio of W vs S to be 2.43:1, thus the Er doping is light. The other signals of C, O and Cu are due to the copper mesh substrate for characterization. To investigate the crystal structure, XRD analysis of the WS$_2$:Er nanosheets is performed. As control sample, WS$_2$ nanosheets are also investigated. Overall, the XRD spectra present peaks attributed to (0 0 2), (0 0 4), (1 0 0), (0 0 6) and (1 1 0) crystal planes as shown in Figure 1d.

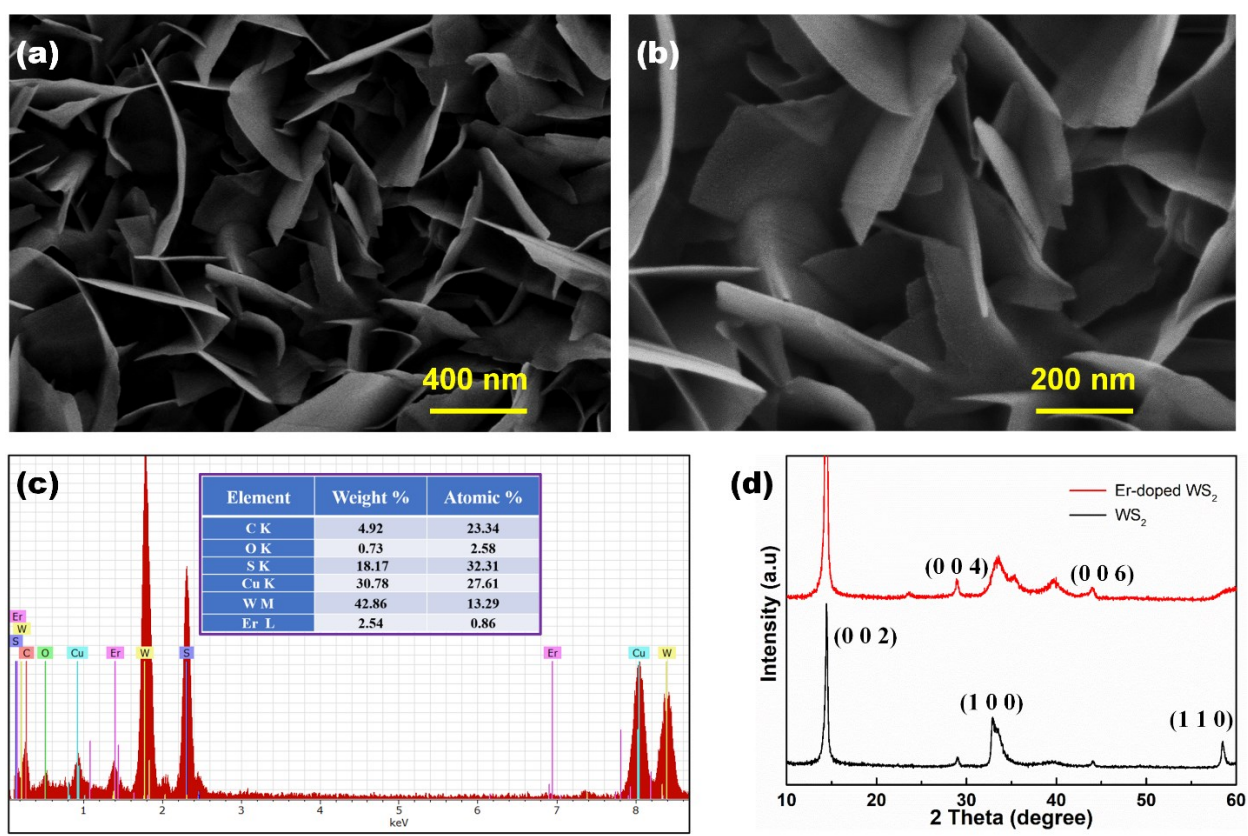

**Figure 1. Morphology, EDS and XRD characterization.** (a) SEM image of WS$_2$:Er nanosheets grown on a Si substrate. (b) Magnified SEM image. (c) EDS spectrum of the WS$_2$:Er nanosheets. The inset shows the elemental contents. (d) XRD patterns of WS$_2$:Er (red line) and pristine WS$_2$ (black line) nanosheets.

Here, the peaks at $2\theta = 14.4°$, $33.4°$ and $58.7°$correspond to the (0 0 2), (1 0 0) and (1 1 0) planes of 2H-WS$_2$ (JCPDF No.84-1398),[47] as can be deduced by comparison with the XRD spectrum of the control sample (black curve in Figure 1d). Upon doping of the WS$_2$ nanosheets with Er, the dominant peaks of (0 0 2) and (1 0 0) do not show evident shift, indicating that the crystal structure is preserved without significant distortion.

TEM and HRTEM measurements are also performed to study the crystal structure of WS$_2$:Er nanosheets. WS$_2$:Er nanosheets grown on the Si substrate are sonicated to obtain single nanosheets, which are deposited on a Cu micro-grid as shown in Figure 2a. The area denoted with dashed line in Figure 2a is magnified to analyze the morphology and structure of the sample (Figure 2b). Figure 2c shows a HRTEM image of the WS$_2$:Er nanosheet to further identify the crystallographic structure, an interplanar spacing of 0.27 nm can be attributed to the lattice plane of (1 0 0) of WS$_2$ with hexagonal phase, which evidences that the incorporation of Er into the WS$_2$ lattices essentially preserves the crystallographic structure. Figure 2d shows a selected area electron diffraction (SAED) pattern of the sample. The two diffraction rings indexed to (100) and (110) planes of 2H-WS$_2$, confirms a polycrystalline nature of the WS$_2$ nanosheets, which is consistent with the XRD results. Figure 2e shows a schematic representation of the sulfurization from cubic phase Er-doped W film into hexagonal phase WS$_2$:Er nanosheets, in which the <0 0 1> side-view crystal lattice of the WS$_2$ with hexagonal phase is presented with Er element substitution for W.

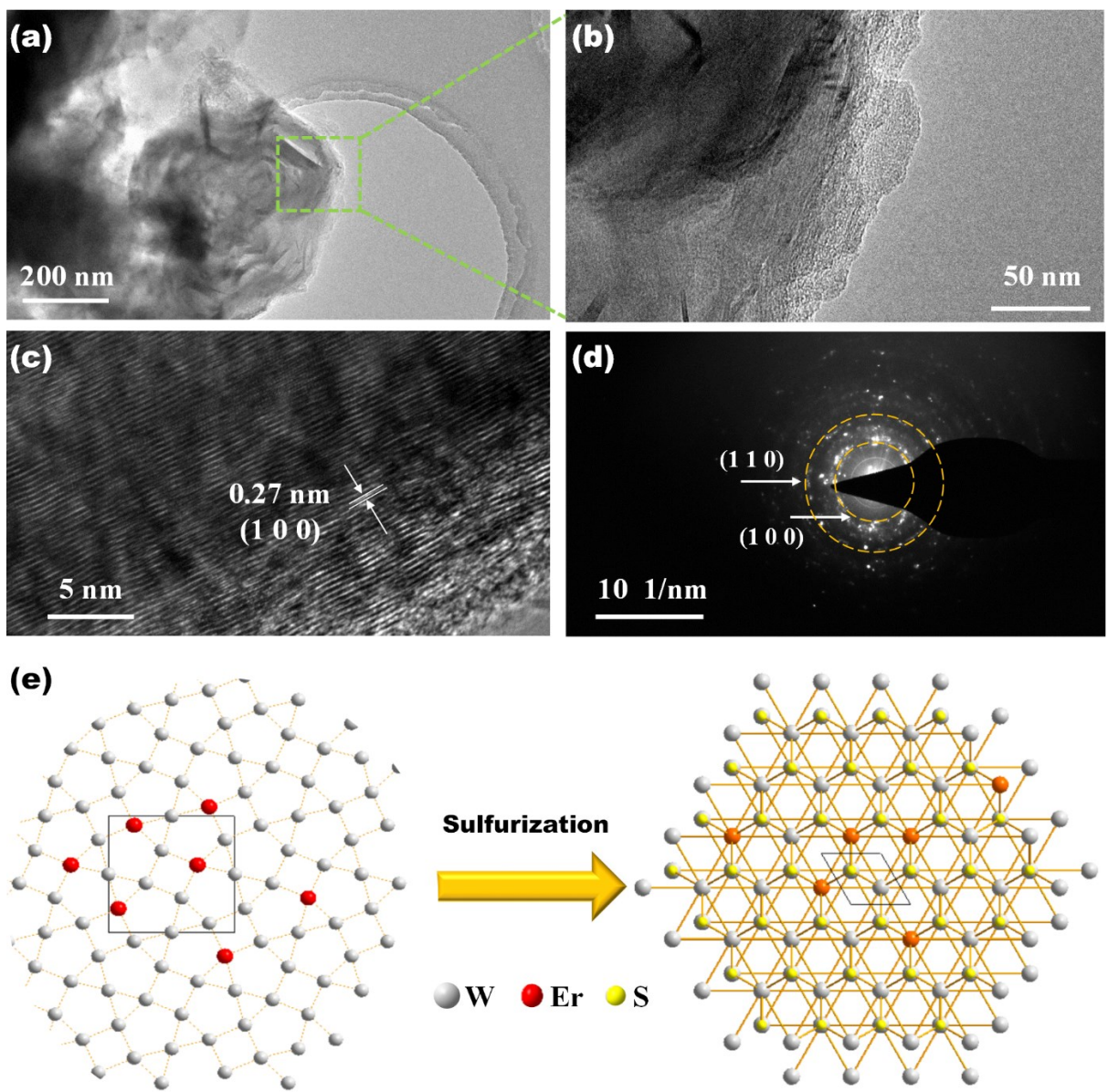

**Figure 2. TEM, HRTEM, SEAD characterization and sulfurization schematics.** (a) Side-view TEM images of the WS$_2$:Er nanosheets. (b) Magnified TEM image of the WS$_2$:Er nanosheet highlighted by a dashed line of (a). (c) HRTEM image of the WS$_2$:Er nanosheet. (d) Corresponding SAED pattern of the WS$_2$:Er nanosheet shown in (c). (e) Schematic presentation of the crystal lattices transformation from Er-doped W film into WS$_2$:Er nanosheet upon sulfurization.

High-resolution XPS spectra are measured to investigate the valence state of the elements. Figure 3a shows the comparisons of XPS results from the pristine $WS_2$ (black line) and $WS_2$:Er (red line). A uniform shift toward lower binding energies of the core-level peaks for W and S in the $WS_2$:Er sample is observed, as compared with those of the pristine $WS_2$, as well as additional features attributed to the core-level peaks of Er. Furthermore, the distinct binding energy peaks corresponding to Er 4d core-levels at 172.4 and 169 eV are found only in the $WS_2$:Er sample as depicted in Figure 3b, which is in accordance with the previously observed XPS features of Er 4d (172.68 and $\approx$168 eV).[45] These results show that the incorporation of Er into the $WS_2$ lattices essentially preserves the $Er^{3+}$ state. Figure 3c shows the S 2p core level spectrum of two peaks centered at 162.8 eV for S $2p_{3/2}$ and 164 eV for S $2p_{1/2}$, which agree with the $S^{2-}$ state. Figure 3d shows the W 4f core-level spectrum, in which two prominent peaks at lower binding energies of 33.1 and 35.3 eV are assigned to W $4f_{7/2}$ and W $4f_{5/2}$, respectively, along with a higher binding energy of 38.8 eV for W $5p_{3/2}$, which are assigned to $WO_3$ and due to oxide contamination. The $WS_2$:Er sample shows relatively high peak intensities at 38.8eV, indicating more oxidization in surface.[52] Overall, the XPS characterizations demonstrate that the Er element is successfully incorporated in $WS_2$ nanosheets.

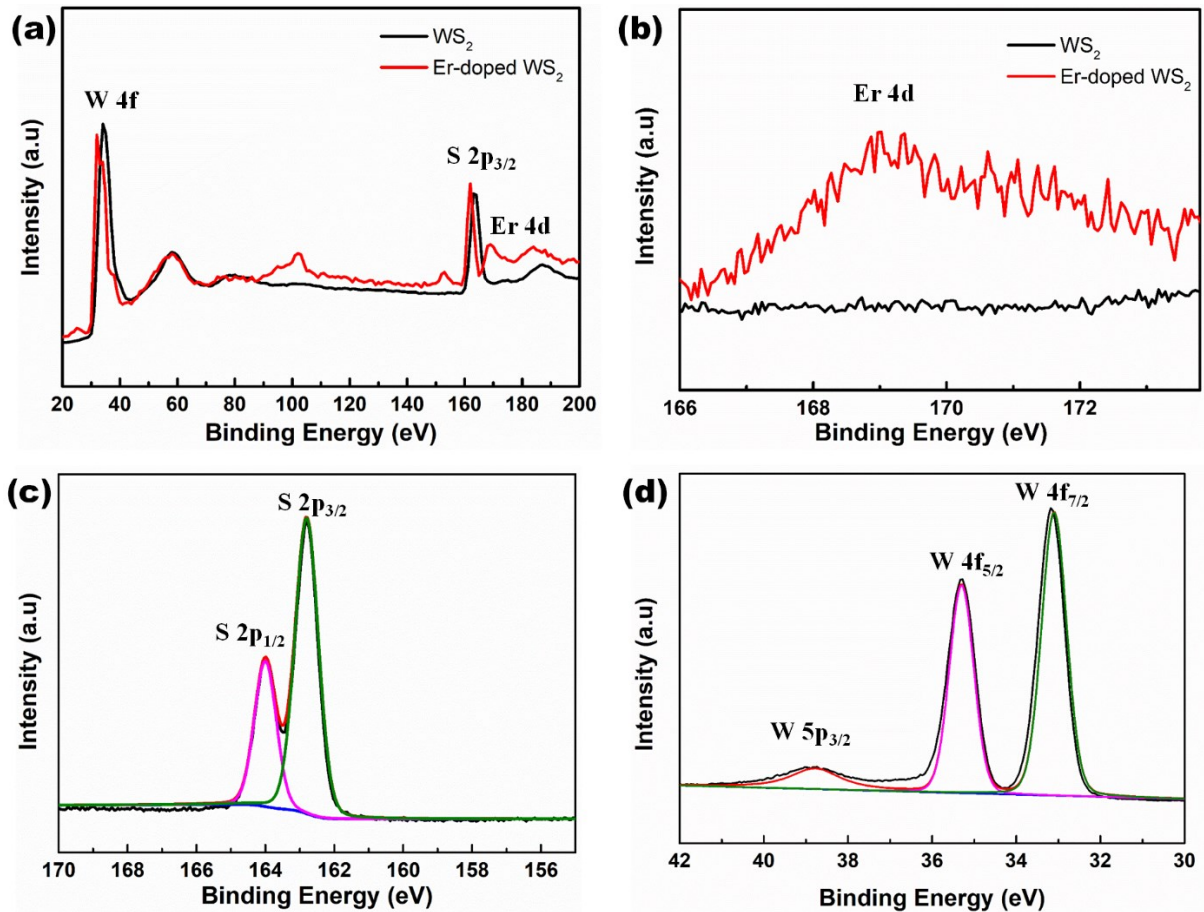

**Figure 3. XPS characterization.** High resolution XPS spectra of (a) full spectrum, (b) Er 4d, (c) S ($2p_{1/2}$ and $2p_{3/2}$), (d) W ($4f_{5/2}$, $4f_{7/2}$ and $5p_{3/2}$) core levels from $WS_2$:Er. As comparison, the results of pristine $WS_2$ are presented in black lines in (a, b).

Raman and PL spectra of the prepared $WS_2$:Er samples are also investigated. Figure 4a shows the main Raman peaks of $WS_2$:Er (red line) under 532 nm laser excitation, which are positioned at 351 $cm^{-1}$ and at 414 $cm^{-1}$ and attributed to in-plane $E_{2g}^1$ vibration and the out-plane $A_{1g}$ vibration of 2H-$WS_2$, respectively. The peaks of $WS_2$:Er are blue-shifted by 2 $cm^{-1}$ compared to the pristine $WS_2$ (black line). The 2 $cm^{-1}$ shift indicates the light doping of Er in $WS_2$ nanosheet.[53] Figure 4b shows PL spectra of the pristine $WS_2$ samples under 532 nm excitation and emission band peaked at 680 nm, with low intensity due to the indirect bandgap of the non-monolayer $WS_2$ nanosheets. Upon Er doping, narrow peaks at 656, 670 and 694 nm are observed, which might be ascribed to free exciton ($X_0$) and trion ($X_S$–singlet; $X_T$–triplet) peaks,[47] as well as the emission from Er due to the radiative decay from $^4I_{9/2}$ to $^4I_{15/2}$ energy level.[54] Moreover, the emission band of $WS_2$:Er features near-infrared peaks at 800 and 860 nm,[48] mainly coming from a one-photon process from the $^4I_{9/2}$ to $^4I_{15/2}$ Er energy level. Under the excitation of a continuous-wave diode laser at 980 nm wavelength, up-conversion PL centered at about 525 nm and 550 nm is found from $WS_2$:Er nanosheets. Emission peaks in this wavelength range are not observed in pristine $WS_2$ nanosheets, as shown in bottom panel of Figure 4b. The PL bands at 525 and 550 nm observed in up-conversion can be attributed to $^2H_{11/2} \rightarrow {}^4I_{15/2}$ and $^4S_{3/2} \rightarrow {}^4I_{15/2}$, respectively, characteristic of Er.[48] To understand down- and up-conversion PL of $WS_2$:Er in-depth, power-dependent PL intensity is performed to elucidate radiative transitions between energy levels of dopant Er in host $WS_2$. Figure 4c shows the power-dependent PL spectra excited by 532-nm laser with power from 2.5 kW.cm$^{-2}$ to 250 kW.cm$^{-2}$, emission peaks at 656, 670 and 694 nm are significantly enhanced upon increasing the excitation power. The emission band of $Er^{3+}$ ions near 800 nm features down-conversion transitions. As shown in inset of Figure 4c, the PL intensity ($I_{PL}$) increases linearly with the pump intensity at 800 nm (similar results obtained for the emission in the interval 640-710 nm, see Figure S2), confirming that the PL from $^4I_{9/2}$ to $^4I_{15/2}$ mainly comes from a one-photon process. To verify the up-conversion transitions, Figure 4d shows the power-dependent PL spectra excited by a continuous-wave diode laser of 980 nm wavelength with power from 50 mW to 170 mW, in which the inset (PL intensity, $I_{PL}$, at 520 nm versus pump power, $P$) shows a power law dependence ($I_{PL} \propto P^n$), with a slope fitted ($n$) to be 2.05±0.10, with indication of an up-conversion process through excited state absorption involving two photons at the pumping wavelength. It should be emphasized that such VIS-NIR down- and up-conversion emissions are simultaneously achieved in the Er-doped $WS_2$, which would bring new opportunities to the development of both fundamental research and applications in NIR photodetectors.

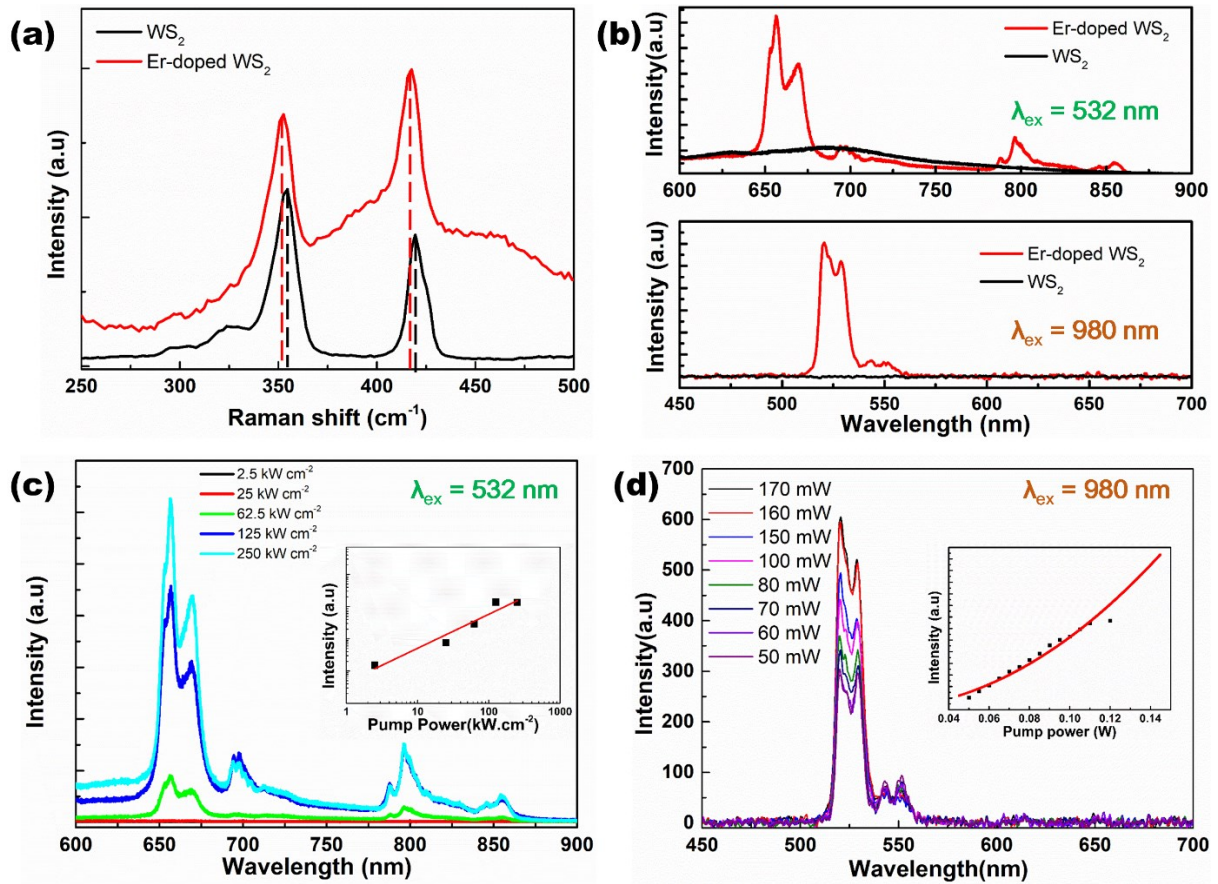

**Figure 4. Raman and PL spectra of WS₂:Er.** (a) Raman spectra of WS₂:Er (red line) and pristine WS₂ (black line). (b) PL spectra of WS₂:Er and pristineWS₂ excited by 532-nm (top panel) and 980-nm (bottom panel) lasers, respectively. (c) Down-conversion PL spectra of WS₂:Er sample pumped with 532-nm laser under different optical power. The inset is the pump power dependence of PL intensity at 800 nm. Red continuous line is a linear fit to the data. (d) Up-conversion PL spectra of WS₂:Er sample pumped with 980-nm laser light at different optical power. The inset is the pump power dependence of PL intensity at 520 nm. The red continuous line is a fit to the data with a power low function, $I_{PL}=A×P^n$ where $A$ and $n$ are fit parameters, by imposing $I_{PL}=0$ for $P=0$.

Figure 5a depicts a scheme of a $WS_2$:Er/Si heterojunction for infrared detection, in which two Ag electrodes are deposited on Si and $WS_2$:Er, respectively. The cross-sectional SEM image of Figure 5b shows a clear interface between Si and $WS_2$:Er nanosheets, and the inset shows a photograph of the prototype device. The current versus voltage ($I$–$V$) characteristics of the $WS_2$:Er/Si heterojunction under different intensity of infrared light are quantitatively analyzed, as well as the responsivity, photoconductive gain and specific detectivity. Figure 5c shows the $I$–$V$ characteristics measured at different infrared intensities under bias from -0.5 V to 0.5 V, signifying the Schottky contact of p-n heterojunction. The current at dark condition is ~50 nA at −0.5 V, indicative of a suppressed dark current which is beneficial for photodetection. We calculate the responsivity ($R$) of the infrared photodetector, $R = (I_{light} − I_{dark})/P_0 S$, where $I_{light}$ and $I_{dark}$ denote the current of device generated under light illumination and in dark , respectively, $P_0$ is the incident power density and $S$ represents the surface area of absorption of infrared light. The values of $R$ at different infrared intensities are plotted in Figure 5d, and display an increase with decreasing infrared intensity with a maximum $R$ value of 39.8 mA/W at an infrared intensity of 4.4 μW. Moreover, the external quantum efficiency ($EQE$) is calculated by $EQE = R \times hc/e\lambda$, in which $R$, $h$, $c$, $e$, and $\lambda$ are the responsivity, the Planck constant, the speed of light, the elementary charge, and the wavelength of incident light, respectively. The $EQE$ value follows the same pattern as the $R$ value variation with light illumination power, as shown in Figure 5d. Following the above results, the detectivity ($D^*$) of photodetector can be determined as $D^* = R/(2eI_{dark}/A)^{1/2}$, where $R$ is the responsivity of the photodetector, $I_{dark}$ refers to the dark current of photodetector, and $A$ is the total cross-sectional area of photodetector. At light power of 4.4 μW, the $D*$ value reaches the highest value of $2.79 \times 10^{10}$ cm·Hz$^{1/2}$/W (Jones), with the lowest value being $1.87 \times 10^{10}$ cm·Hz$^{1/2}$/W (Jones) at 18.0 μW, as shown in Figure 5e. For comparison, a photocurrent of 10 nA and a responsivity of 0.8 mA/W are obtained for a device without Er-doping (see Figure S3) at an infrared intensity of 4.4 μW with bias voltage of -0.5V. The high value of responsivity and detectivity is attributed to the effect of the p-n junction inner field formed between p-type Si and n-type Er-doped WS$_2$, suppressing the dark current of device, along with infrared photon absorption enhancement, and defect states in WS$_2$ suppressed by Er doping. Moreover, as key parameter of a photodetector, the response speed represents its ability to follow a fastly-varied light signal. The time response of rise and fall signals (~170/155 ms) are obtained, which is ascribed to the interstate trapping in WS$_2$ for photo-induced electron-holes separation and escaping.[47] Overall, the above results confirm that the synergetic effect of the Er-doping and the heterojunction architecture improves the infrared photoresponse of the photodetector. Thus, our Er-doped-WS$_2$/Si heterojunction photodetector can be sensitive to weak infrared light signals even at the bias of only −0.5 V.

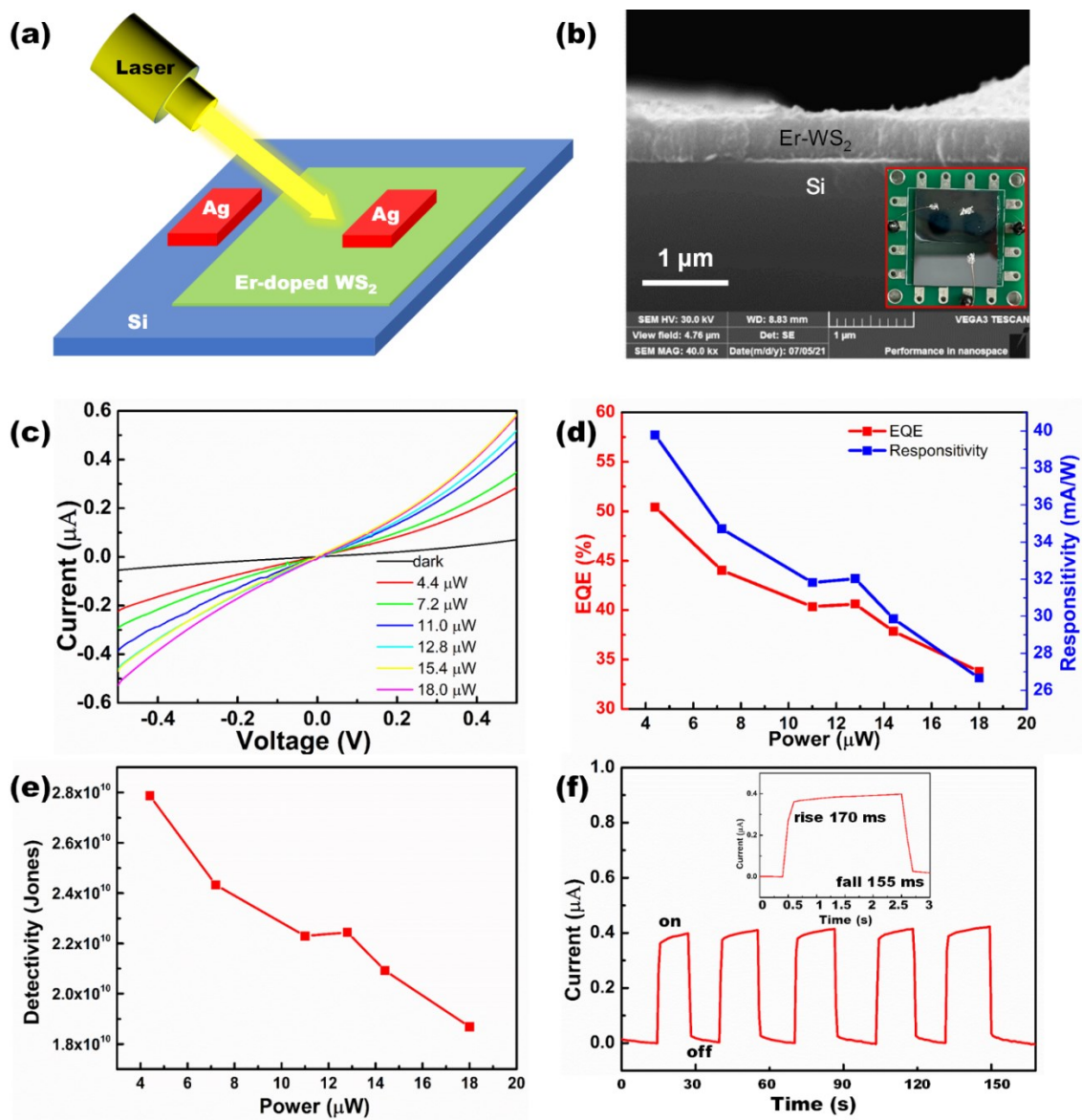

**Figure 5. Photodetection performance of WS₂:Er/Si Heterojunction.** (a) Schematics of the infrared photodetector. (b) Cross-sectional SEM image of Er-WS₂/Si heterojunction. Inset: photograph of the prototype device. (c) Current-voltage curves of the infrared photodetector under 980-nm illumination with powers from 0 to 18.0 μW. (d, e) Responsivity, *EQE* and detectivity as a function of illumination power at bias voltage of −0.5V. (f) Time dependent response of the photodetector with 980 nm On/Off illumination at -0.5 V bias. The inset shows the rise and fall time of 170 and 155 ms, respectively.

The high value of photoresponse of the device can be understood by the energy band diagram in Figure 6a. Since pristine $WS_2$ nanosheest has indirect bandgap with lower photon absorption, S vacancies are new defect sites in the basal plane since gap states around the Fermi level, which allow direct photon absorption more easily and the induced carriers to be collected into electrodes more easily for detection. In our work, S-vacancies in $WS_2$ in the basal plane result in infrared photon absorption enhancement for alleviating the photoelectrons recombination in the heterostructure interface, which would support our experimental results more strongly. Type-n $WS_2$ film contacts the type-p Si substrate to form a p-n vertical heterojunction with lower suppressed dark current. When 980 nm light illuminates the device, the electron of valence state of Si is excited to conductance state, whereas $WS_2$ only absorbs some photons by defects states. Finally, the photo-induced carriers are swept by the inner field to form the detection current injected into Ag electrodes. When $Er^{3+}$ doping is introduced into the host $WS_2$, the above-mentioned optical absorbance mechanism is observed together with energy cross transfer between $Er^{3+}$ and $WS_2$. The excited state absorption of 980-nm light leading to up-conversion mechanisms from $Er^{3+}$ boost the electron from ground state $^4I_{15/2}$ into high energy states, such as $^2H_{11/2}$ and $^4S_{3/2}$, resulting in the 520-nm and 550-nm light emission. Meanwhile, one-photon absorption excites ground state $^4I_{15/2}$ into high energy state $^4I_{11/2}$, the photon lifetime is short and dissipates into more stable state $^4I_{13/2}$ (1550-nm light radiation). The energy of green light emitted by $Er^{3+}$ can be transferred to $WS_2$ host, resulting in enhanced photon-induced electron-hole pairs, which yields more photocurrent in device based on Er-doped $WS_2$. To further confirm the above discussed gain mechanism and especially the p-n junction suppressing dark current, a device simulation is carried out. The optical field energy density distribution under 980 nm light illumination is presented in Figure 6b. The light passes from the device top area without electrodes blocking, and a periodic distribution of the electric field with strong and weak pattern inside the device due to light wave characteristic. The optical field energy density distribution is indicative of optical absorption capability in different device areas, and higher energy distribution can lead to more photo-induced electron-hole pairs. Herein, only the n-WS$_2$/p-Si heterojunction devices are simulated to understand current density field distributions under dark condition (Figure 6c). The current density field is mainly distributed on the device electrodes with lower current density. This result is in agreement with the suppressed dark current from the p-n junction. The light-induced current is obviously manifested in the current density field distributions of Figure 6d (about four orders of magnitude higher than that of Figure 6c). Moreover, Figure 6e shows the $I$–$V$ characteristic of n-WS$_2$/p-Si device with same parameters as Figure 6 (c-e), in both dark and light conditions. The simulation is performed under 980-nm light illumination with 0.01 mW. We find that, in the dark condition, the device has higher dark current (~ 2.8 nA at -0.5 V applied voltage), whereas a current of ~280 nA is obtained at the same applied voltage under light illumination, with responsivity ~28 mA/W. Upon considering the Er$^{3+}$ doping-related up-conversion mechanism, the total responsivity would be larger due to the higher photo-induced electron-hole pairs generation. Overall, the Er$^{3+}$ doping in host WS$_2$ layers could effectively enhance photon-induced electron-hole pairs contributing to current, and then suppress the dark current of the heterojunction photodetector by reducing minority carrier transportation between WS$_2$ and Si, thus further improving the performance of the device.

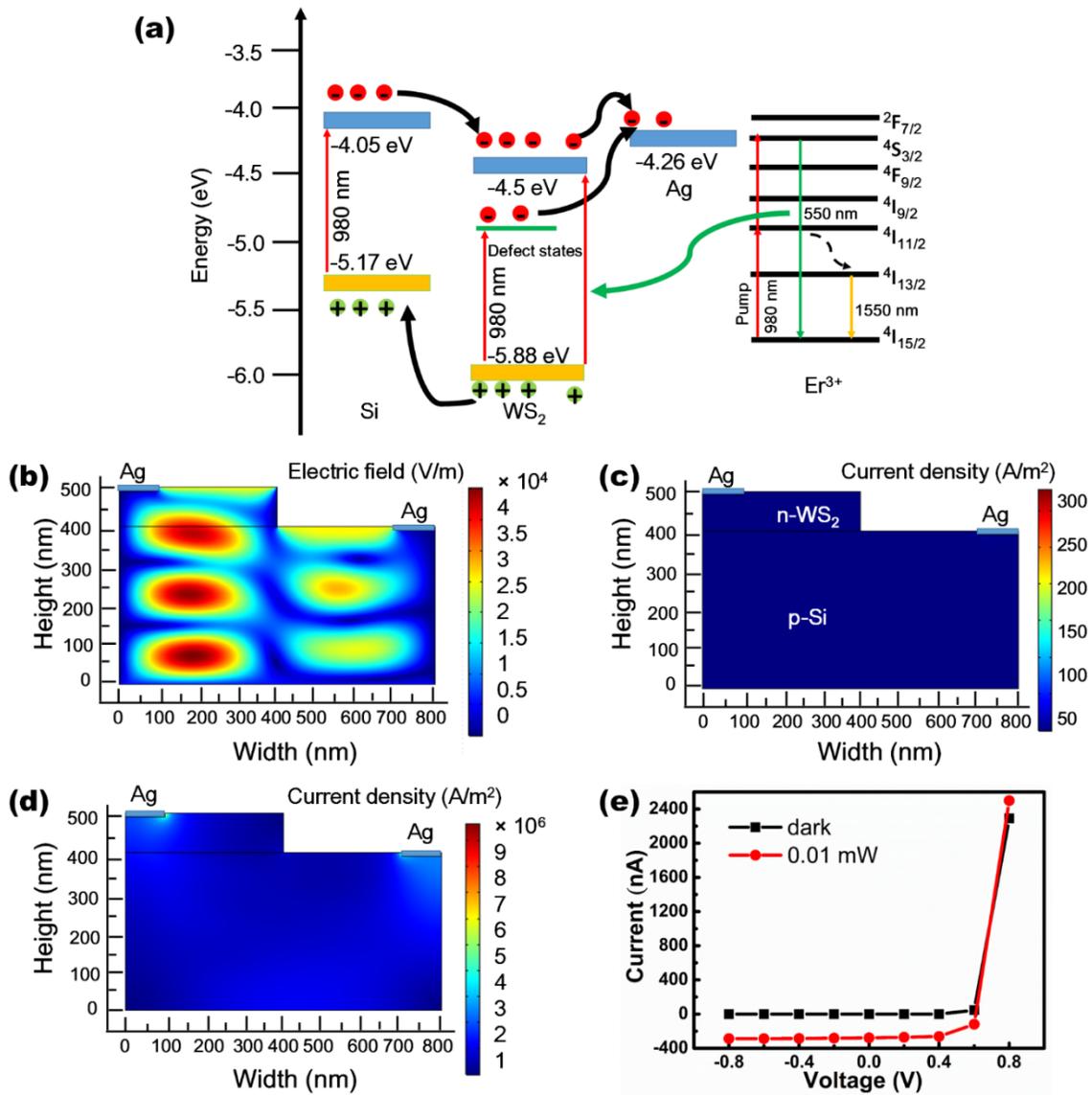

**Figure 6. Performance analysis for photodetection mechanism.** (a) Schematic energy level diagram of the WS$_2$:Er/Si heterojunction, showing energy transitions for up-conversion due to Er$^{3+}$ dopant. (b) Simulated optical field distribution of the WS2:Er/Si heterojunction device. (c, d) Current density field distribution of the WS$_2$:Er/Si heterojunction device in dark condition (c) and under light illumination (d) at 980 nm with power of 0.01 mW. (e) Simulated $I-V$ curves of the WS$_2$:Er/Si heterojunction device in the dark and under light illumination, from the same device configuration as in (b-d).

## 3. CONCLUSION

WS$_2$:Er/Si heterojunction devices for infrared detection are successfully fabricated by a simple two-step fabrication involving co-sputtering of a W:Er film and sulfurization into WS$_2$:Er on p-type Si substrate. WS$_2$:Er nanosheets achieve simultaneous up-conversion and down-conversion PL ranging from visible to the near-infrared region suitable for light detection. Infrared photodetectors based on WS$_2$:Er/Si vertical heterojunctions with suppressed dark current are achieved with a responsivity of 39.8 mA/W at a weak 980-nm light with power of 4.4 µW and a detectivity of $2.79 \times 10^{10}$ Jones. The photodetection performance is elucidated by the energy band diagram of energy cross transition related to up-conversion phenomenon, as well as numerical simulations. This Er-doped WS$_2$ can be utilized as a good candidate for high-performance photodetectors in the infrared regions, and for several future optoelectronic platforms.

## ACKNOWLEDGEMENTS

The work was supported by National Natural Science Foundation of China (Grant No. 11804120), and the Science Foundation of Guangzhou Program (No. 2023A03J00229), Outstanding Youth Cultivation Program of Huizhou University (HZU202017), The Professorial and Doctoral Scientific Research Foundation of Huizhou University (Grants No. 2020JB043) and The Program for Innovative Research Team of Guangdong Province & Huizhou University (IRTHZU). A. C. acknowledges the funding from the European Research Council under the European Union's Horizon 2020 Research and Innovation Program (Grant Agreement no. 682157, "xPRINT"). D. P. acknowledges the funding from the Italian Minister of University and Research PRIN 2017PHRM8X project ("3D-Phys"), and the PRA_2018_34 ("ANISE") project from the University of Pisa. X. Y. is grateful to Prof. Baojun Li, Long Wen, and Qin Chen from Jinan University. H. R. is grateful to MS. Zhiwei Sun and Bojun Chen for experimental assistance.

**Supporting Information**

**Erbium-doped WS₂ with Down- and Up-Conversion Photoluminescence Integrated on Silicon for Heterojunction Infrared Photodetection**

*Qiuguo Li,* [1] *Hao Rao,* [2] *Haijuan Mei,* [1] *Zhengting Zhao,* [1] *Weiping Gong,\*,* [1] *Andrea Camposeo,* [3] *Dario Pisignano,* [3, 4] *and Xianguang Yang\*,* [2]

[1]Guangdong Provincial Key Laboratory of Electronic Functional Materials and Devices，Huizhou University, Huizhou 516001, Guangdong, China

[2]Institute of Nanophotonics, Jinan University, Guangzhou 511443, China

[3]NEST, Istituto Nanoscienze-CNR and Scuola Normale Superiore, Piazza S. Silvestro 12, I-56127 Pisa, Italy

[4]Dipartimento di Fisica, Università di Pisa, Largo B. Pontecorvo 3, I-56127 Pisa, Italy

*E-mail: xianguang@jnu.edu.cn and gwp@hzu.edu.cn

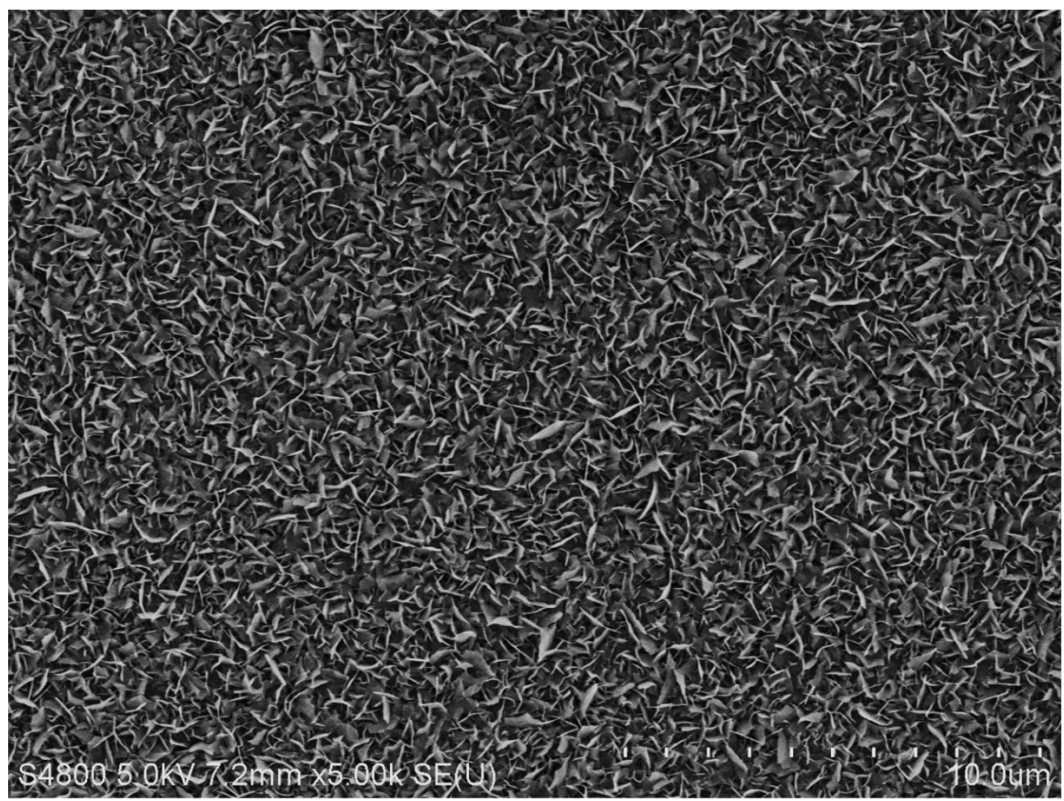

**Figure S1**. SEM micrograph for a large area of uniform WS$_2$:Er nanosheets grown on a Si substrate.

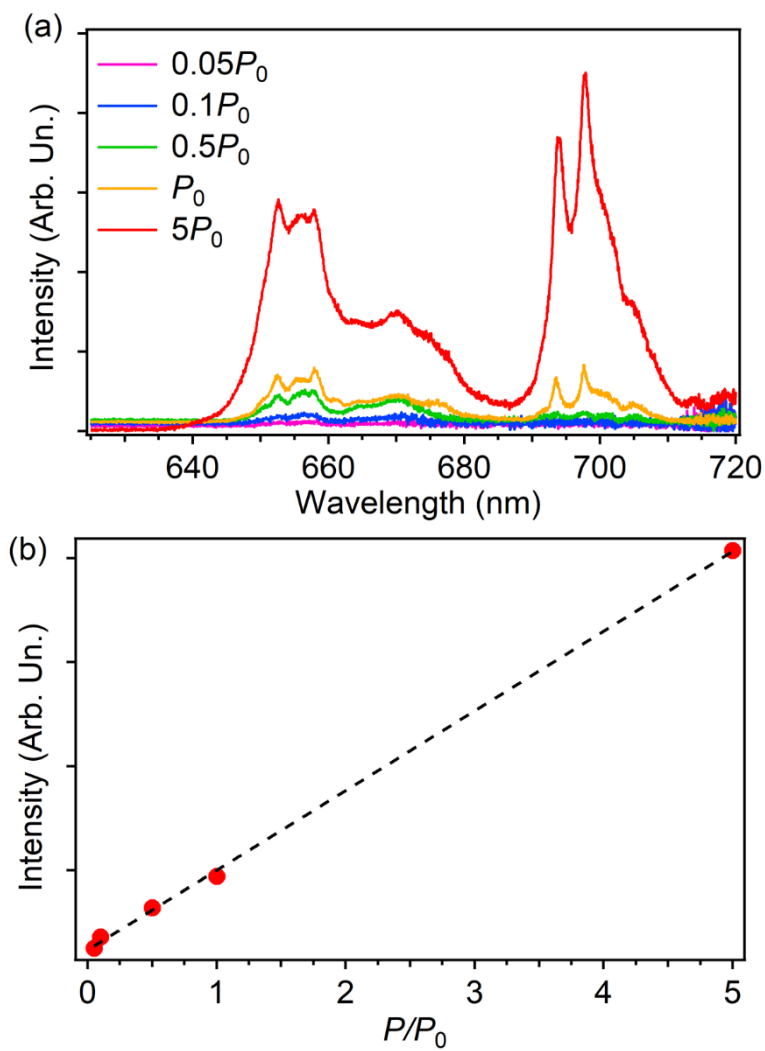

**Figure S2**. (a) Down-conversion PL spectra of $WS_2$:Er sample pumped with 532-nm laser under different optical power, in which the emission is in the interval 640-710 nm. (b) Pump power dependence of PL intensity at 656 nm with red dot denotation. The black dashed line is a fit to the data with a power low function, $I_{PL} = A \times P^n$, where $A$ and $n$ are fit parameters, by imposing $I_{PL} = 0$ for $P = 0$.

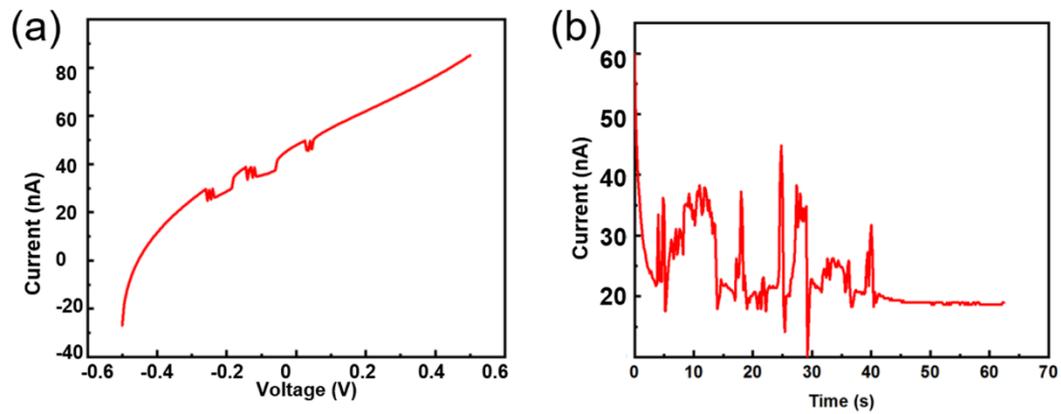

**Figure S3**. (a) I-V characteristic curve of a device with WS$_2$:Si junction without erbium doping under dark conditions. (b) The photoelectric response curve of a device with a WS$_2$:Si junction without erbium doping under the irradiation of a 980 nm laser with infrared intensity of 4.4 µW at bias voltage of -0.5V.